\documentclass[useAMS,usenatbib,usegraphicx]{mn2e}
\usepackage{amssymb,amsmath, color, soul, ulem,graphicx,commath}

\newcommand{\plm}{P_{\rm lm}}
\newcommand{\lm}{_{\rm lm}}
\newcommand{\tp}{(\theta, \varphi)}
\newcommand{\plmprime}{P_{\rm l' m'}}
\newcommand{\dplm}{\frac{\dif \plm}{\dif \theta}}
\newcommand{\lmp}{_{\rm l'm'}}

\title[The magnetic field vector of the sun-as-a-star]{The magnetic field vector of the sun-as-a-star}
\author[A.~A.~Vidotto]{A.~A.~Vidotto$^{1,2}$\thanks{E-mail: Aline.Vidotto@tcd.ie}
\\ 
$^{1}$School of Physics, Trinity College Dublin, Dublin-2, Ireland\\
$^2$University of Geneva, Chemin des Maillettes 51, Versoix, CH-1290, Switzerland
}

\date{Accepted XXX. Received YYY; in original form ZZZ}

\pubyear{2016}

\begin{document}
\label{firstpage}
\pagerange{\pageref{firstpage}--\pageref{lastpage}}
\maketitle

\begin{abstract}
Direct comparison between stellar and solar magnetic maps are hampered by their dramatic differences in resolution. 
Here, we present a method to filter out the small-scale component of vector fields, in such a way that comparison between solar and stellar (large-scale) magnetic field vector maps can be directly made. Our approach extends the technique widely used to decompose the radial component of the solar magnetic field to the azimuthal and meridional components as well. 
For that, we self-consistently decompose the three-components of the vector field using spherical harmonics of different $l$ degrees. By retaining the low $l$ degrees in the decomposition, we are able to calculate the large-scale magnetic field vector. 
Using a synoptic map of the solar vector field at Carrington Rotation CR2109, we derive the solar magnetic field vector at a similar resolution level as that from stellar magnetic images. We demonstrate that the large-scale field of the Sun is not purely radial, as often assumed -- at CR2109, $83\%$ of the magnetic energy is in the radial component, while $10\%$ is in the azimuthal and $7\%$ is in the meridional components. By separating the vector field into poloidal and toroidal components, we show that the solar magnetic energy at CR2109 is mainly ($>90\%$) poloidal. Our description is entirely consistent with the description adopted in several stellar studies. Our formalism can also be used to confront synoptic maps synthesised in numerical simulations of dynamo and magnetic flux transport studies to those derived from stellar observations. 
\end{abstract}

\section{Introduction}
\subsection{Small-scale structures of magnetic fields}
Studies of solar magnetism have provided us with fantastic spatial (i.e., enabling us to resolve small-scale structure of the solar magnetic fields) and temporal (with cadences reaching less than a minute) resolutions. Studying magnetism in stars, although more observationally challenging, is equally rewarding.  In particular, imaging the magnetic field of solar-type stars, despite being less detailed, allow us to put the Sun in a much more general context. 

Solar observations have revealed a multitude of details of the solar photospheric magnetic field. Although observations have not yet fully resolved all the solar magnetic structures, these structures are seen in a wide dynamical range: small-scale structures organised into ephemeral regions, network features and internetwork features (as small as a few $1000$~km, \citealt{2011SoPh..272...29M})  form the magnetic carpet of the quiet Sun. As the Sun moves towards increasing activity in its cycle, sunspots start to show up more frequently in the solar photosphere. Sunsposts appear in pairs of opposite magnetic polarity and group together, forming active regions that are then distributed at low latitudes over the solar surface. The stellar counterparts to the solar small-sized magnetic features are currently not resolved in images of stellar magnetism.   

The main technique used to image stellar surface magnetic field is the Zeeman Doppler Imaging (ZDI) technique. This technique consists of analysing a series of circularly polarised spectra (Stokes V) to recover information about the large-scale magnetic topology, including its intensity, orientation and how this field is distributed over the stellar surface \citep{1997A&A...326.1135D}. In practice,  the star is observed during several rotation cycles. Then, the time-series of high-resolution Stokes V profiles is inverted into a surface magnetic map. ZDI studies have demonstrated that cool dwarf stars harbour at their surfaces large-scale magnetic fields with a wide variety of intensities and topologies  \citep[e.g.][]{2006Sci...311..633D,2006MNRAS.370..468M,2007MNRAS.377.1488H,2008MNRAS.388...80P,2009MNRAS.398.1383F,2010MNRAS.407.2269M,2016MNRAS.457..580F}.

The ZDI technique is able to reconstruct the topology (polarity and orientation) of the stellar magnetic field through the stellar surface. However, these images have much lower resolution than solar magnetic field maps due to several factors, such as the temporal sampling of the observations, their achieved signal-to-noise ratio, the intrinsic width of the (local) stellar line profile and the resolution of observations. Regarding the latter, magnetic fields of different polarities within an angular resolution element cancel each other out.\footnote{Magnetic elements of different polarities can be detected only if their spectral line profiles are significantly shifted from each other to prevent their complete overlap and cancellation. In its turn, this requires a significant Doppler shift between the line profiles, which can be achieved only for large spatial separations between two flux elements.} As a result, small-scale structures cannot be seen by ZDI, which instead is able to map only the large-scale magnetic field \citep{2010MNRAS.404..101J, 2011MNRAS.410.2472A, 2014MNRAS.439.2122L}. 

\subsection{Synoptic maps of solar and stellar magnetic fields}
\begin{figure*}
\includegraphics[width=0.99\textwidth]{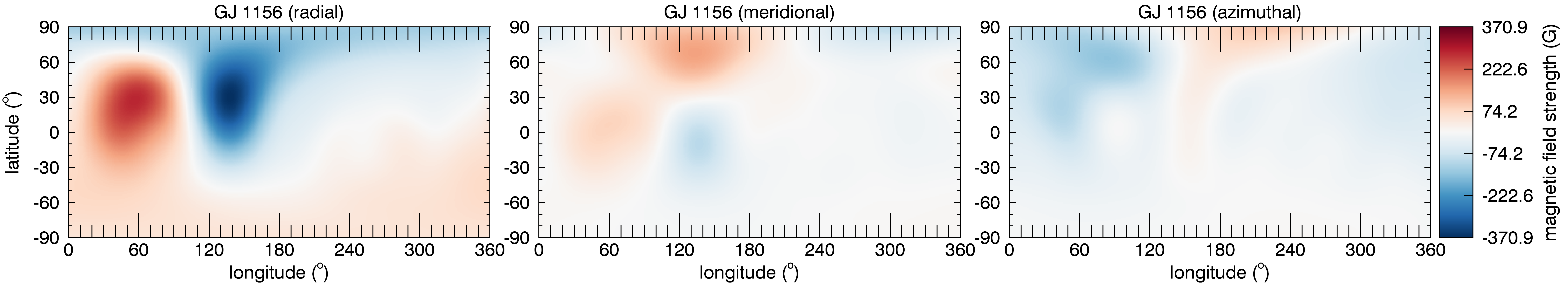}
\caption{The magnetic field of the star GJ1156 observed in 2008, reconstructed \citep{2010MNRAS.407.2269M} using the ZDI technique. From left to right: the radial, meridional (North-South) and azimuthal (East-West) components. This synoptic map is expressed in terms of spherical harmonics up to the maximum degree $l_{\rm max}=6$. \label{fig.magmap}}
\end{figure*}

Synoptic maps of vector magnetic fields are one of the outputs of the ZDI technique. Through its spectropolarimetric monitoring during several stellar rotation cycles, ZDI can reconstruct the surface large-scale vector field in its three components: radial $B_r$, meridional $B_\theta$ (North-South) and azimuthal $B_\varphi$ (East-West) components (Figure \ref{fig.magmap}). 
%

Surprisingly, and in spite of the significant advances in solar observations for over half a decade, only the strength and the line-of-sight (LOS) component of the solar magnetic field have been systematically measured. Vector magnetic fields at the photosphere are available, but mainly for localised small areas of the solar surface (e.g., active regions \citealt{2006ApJ...646L..85Z,2014SoPh..289.3549B}). Full-disk vector magnetograms, which are required to create synoptic maps of the vector solar field, have just recently started to be measured with the Helioseismic and Magnetic Imager (HMI, \citealt{2012SoPh..275....3P,2014SoPh..289.3483H}) at the Solar Dynamics Observatory and by the Synoptic Optical Long-term Investigations of the Sun (SOLIS \citealt{2003SPIE.4853..194K}) at the National Solar Observatory. The global capabilities of these instruments will provide us a unique opportunity to observe the large-scale spatial distribution of vector magnetic fields across the solar surface \citep{2012LRSP....9....6M}. 

Although synoptic maps of the stellar vector field exist for a couple of decades and have been derived for hundreds of systems \citep[e.g.][]{1997MNRAS.291....1D,1999MNRAS.302..437D,2006MNRAS.370..629D,2009ARA&A..47..333D}, there are significantly fewer synoptic maps of the vector field of the Sun. Earlier synoptic maps of the  vector solar field were created on the basis of the rotational modulation of LOS magnetograms \citep[][also called ``pseudo-vector reconstruction method'']{2000ApJ...528..999P, 2010ApJ...720..632W}. With this method, these authors assumed that the field on large spatial scales does not change over several days. Because of this assumption, these early synoptic maps reconstructed only large-scale magnetic fields of the Sun. More recently,  \citet{2013ApJ...772...52G} used daily full-disk vector magnetograms from the SOLIS/VSM spectrograph to reconstruct the vector solar field of a series of 23 Carrington Rotations. This is a significant advance compared to the large majority of synoptic charts of the solar magnetic field, which are restricted to the LOS component of the solar magnetic field, used to derive $B_r$. An important conclusion reached by \citet{2013ApJ...772...52G} is that the solar magnetic field is not normal (i.e., purely radial) to the surface at the photospheric level. This is somewhat disturbing, as the assumption that the solar magnetic fields are purely radial are used in several studies.

With the new existing capabilities in producing vector synoptic fields, we can now start to investigate the behaviour of the meridional $B_\theta$ and azimuthal $B_\varphi$ components of the solar magnetic field and investigate how the three magnetic field components in the Sun compare to the ones derived in stellar observations (e.g., Figure \ref{fig.magmap}). However, direct comparison between stellar  and solar magnetic field topologies are hampered by their dramatic differences in resolutions. To provide a direct comparison between solar and stellar magnetic field synoptic maps, it is first required to `filter out' the small-scale field (i.e., the finely resolved magnetic features) of the solar observations, since this component is currently not accessible to ZDI studies. After this, we are left with only the solar large-scale field, whose spatial scale is comparable to those achieved by stellar observations.

One way to filter out the small-scale structure of the solar field is to express the solar magnetic field as spherical harmonics of different $l$ degrees, up to a maximum degree $l_{\rm max}$. The smallest $l$ degrees represent the largest-scale components, e.g., $l=1$ for the dipole, $l=2$ for the quadrupole, $l=3$ for the octupole and so on. By retaining only the components of low $l$ degrees in the spherical harmonics decomposition, we are then able to calculate the large-scale component of the magnetic map. This approach has been extensively applied to the LOS and radial components of the solar field \citep[e.g.][]{2003JGRA..108.1035S,2012ApJ...757...96D}. The choice of $l_{\rm max}$ is  related to the spatial resolution of the maps. In high-resolution solar synoptic maps, $l_{\rm max} =192$ \citep{2012ApJ...757...96D}, while in stellar synoptic maps, $l_{\rm max} \lesssim 10$ \citep{2014MNRAS.437.3202J}, quantitatively demonstrating the large-scale nature of the currently available ZDI measurements. In the map shown in Figure \ref{fig.magmap}, for example, $l_{\rm max}=6$ \citep{2010MNRAS.407.2269M}. 

To the best of our knowledge, a mathematical description of the solar magnetic field using spherical harmonics is currently only available for its radial component, limiting us to draw analogies between solar and stellar fields {\it only} for the radial large-scale magnetic field component. In this paper, we extend the currently available mathematical description of $B_r$ to $B_\theta$ and $B_\varphi$ components. Our description is entirely consistent with the description adopted in several ZDI studies \citep[e.g.][]{2006MNRAS.370..629D,2015MNRAS.453.3706D,2010MNRAS.407.2269M,2012MNRAS.423.1006F,2012A&A...540A.138M} making it straightforward to compare the large-scale magnetic field vector of the Sun at the same resolution of stellar studies. In practice, with the set of equations we derive in the present study, one will be able to derive the spherical harmonics coefficients from a synoptic map of the vector field (either from solar observations or synthesised in numerical simulations of dynamo and magnetic flux transport studies), filter out the large $l$-degrees (i.e., keeping the degrees that represent the large-scale field) and reconstruct only the large-scale field component. Alternatively, by filtering out the small $l$-degrees (i.e., keeping the degrees that represent the small-scale field), one can also study the small-scale field distribution. 

This paper is divided as follows: Section \ref{sec.mathematical} presents the method we use to decompose the magnetic field vector using spherical harmonics and its inversion. Section \ref{sec.application} illustrates applications of our method and Section \ref{sec.conclusions} presents a discussion and our conclusions. 

\section{Magnetic field decomposition using vector spherical harmonics}\label{sec.mathematical}
\subsection{The mathematical description used by ZDI}
The ZDI technique consists of reconstructing the stellar surface magnetic field based on a series of circularly polarised spectra \citep{1997A&A...326.1135D}. Several implementations of the technique exist \citep[e.g.,][]{1997MNRAS.291....1D,2002ApJ...575.1078H,2006MNRAS.370..629D,2016A&A...586A..30K}. In this work, we follow the implementation from \citet{2006MNRAS.370..629D}.
In this implementation, ZDI solves for the radial $B_r$, meridional $B_\theta$ and azimuthal $B_\varphi$ components of the stellar magnetic field, expressed in terms of spherical harmonics and their colatitude-derivatives\footnote{Similar sets of equations are also used by \citet{2014A&A...565A..83K}. Note that only the real part of equations (\ref{eq.br}) to (\ref{eq.bphi}) are used. We also note that Equations (\ref{eq.br}) and (\ref{eq.btheta}) have different signs as those in \citet{2006MNRAS.370..629D}, because of different coordinate systems adopted. In the present paper, radial field points outwards, the meridional ($\theta$) component increases from North to South poles and the azimuthal ($\varphi$) component increases in the direction of rotation (increasing longitude or decreasing rotational phase).}
\begin{equation}\label{eq.br}
B_r \tp =    \sum\lm \alpha\lm Y\lm \tp \, ,
\end{equation}
\begin{equation}\label{eq.btheta}
B_\theta \tp =   \sum\lm \beta\lm Z\lm\tp + \gamma\lm X\lm \tp  \, ,
\end{equation}
\begin{equation}\label{eq.bphi}
B_\varphi \tp =- \sum\lm \beta\lm X\lm\tp - \gamma\lm Z\lm \tp  \, , 
\end{equation}
where 
\begin{equation}\label{eq.ylm}
Y\lm\tp = P\lm (\cos \theta) e^{im\varphi} \, ,\end{equation}
\begin{equation}\label{eq.xlm}
X\lm\tp = \frac{1}{(l+1) \sin\theta} \frac{\partial Y\lm \tp}{\partial \varphi} = \frac{ im \plm  e^{im\varphi}}{(l+1) \sin\theta}\, ,
\end{equation}
\begin{equation}\label{eq.zlm}
Z\lm\tp = \frac{1}{l+1} \frac{\partial Y\lm \tp}{\partial \theta} = \frac{1}{l+1} \frac{\dif \plm}{\dif \theta} e^{im\varphi}   \, .
\end{equation}
$ \plm \equiv \plm (\cos \theta)$ is  the associated Legendre polynomial of degree $l$ and order $m$. 
$\alpha\lm$, $\beta\lm$, $\gamma\lm$ are the coefficients that provide the best fit to the spectropolarimetric data and are such that Equations (\ref{eq.br}) to (\ref{eq.bphi}) obey the solenoidal constraint on the magnetic field ($\boldsymbol{\nabla}\cdot \mathbf{B}=0$).  
The sums should be performed over $1\leq l \leq l_{\rm max}$  and  $|m| \leq l$, where $l_{\rm max}$ is the maximum degree of the spherical harmonic decomposition.  
Alternatively, if one prefers to consider only positive $m$ values ($0\leq m \leq l$), as we do in the present paper, then a factor $(2-\delta_{m,0})$ should be included in the sums, where $\delta_{m,0}=1$ for $m=0$ and $\delta_{m,0}=0$ for $m\ne 0$.\footnote{The inclusion of the factor $(2-\delta_{m,0})$ in the sums running only through positive values of $m$ is possible because the spherical harmonics with negative values of $m$ are related to the positive $m$ value components. This is demonstrated in, e.g., \citet{2012PhDT........75J}. We take this approach because it speeds up calculation, as we do not have to compute the sums for negative $m$ values.} Taking Eq.~(\ref{eq.br}) as an example, we have that $\sum\lm \alpha\lm Y\lm = \sum_{l=1}^{l=l_{\rm max}} \sum_{m=-l}^{m=l} \alpha\lm Y\lm = \sum_{l=1}^{l=l_{\rm max}} \sum_{m=0}^{m=l} (2-\delta_{m,0}) \alpha\lm Y\lm$. We further note that \citet[][see also \citealt{2014A&A...565A..83K}]{2006MNRAS.370..629D} define $Y\lm\tp = c\lm P\lm (\cos \theta) e^{im\varphi} $, where $c\lm$ is a normalisation constant 
\begin{equation}
c\lm = \sqrt{\frac{2l+1}{4\pi}\frac{(l-m)!}{(l+m)!}}.
\end{equation}
In our Equations (\ref{eq.ylm}) to (\ref{eq.zlm}), $c\lm$  is absorbed in the definition of $P\lm (\cos\theta)$. 

\subsection{Inversion of the magnetic field equations}  
The large majority of solar synoptic maps provide only the LOS magnetic field component of the solar photosphere. This component is then transformed in a radial component $B_r$. Several solar observatories provide the spherical harmonic coefficients $\alpha\lm$, so that to obtain the radial distribution of solar magnetic field $B_r \tp$, one should use Equation (\ref{eq.br}). Alternatively, when solar observatories provide the distribution of  $B_r \tp$, usually in the form of a bi-dimensional array stored in a fits file format, one can invert Equation (\ref{eq.br}) to compute $\alpha\lm$. This method has been widely used and is very powerful to extract the large-scale component of the solar radial magnetic field \citep{2012ApJ...757...96D}, including in studies of potential field extrapolation \citep[e.g.][]{1969SoPh....9..131A,1977SoPh...51..345A,1992ApJ...392..310W,2000GeoRL..27..505W,2003JGRA..108.1035S,2013ApJ...768..162P}. It is straightforward to invert Equation (\ref{eq.br}), using the mathematical properties of the associated Legendre polynomials \citep[e.g.][]{1969SoPh....9..131A,1977SoPh...51..345A}.

However, since most solar synoptic maps have only provided the radial component of the solar magnetic field, the inversion of Equations (\ref{eq.btheta}) and (\ref{eq.bphi}) to obtain the coefficients $\beta\lm$ and $\gamma\lm$ has not been derived in the literature to the best of our knowledge. In this section, we present a derivation of the coefficients $\alpha\lm$, $\beta\lm$ and $\gamma\lm$, from a bi-dimensional distribution of the surface $B_r$, $B_\theta$ and $B_\varphi$. Before doing that, we present next a couple of orthogonal properties of vector spherical harmonics that will be used in our derivations.

\subsubsection{Orthogonal properties of the vector spherical harmonics}
The vector spherical harmonics obey  the following orthogonal properties \citep{1985EJPh....6..287B, 1991EJPh...12..184C}:
\begin{equation}\label{eq.prop1}
\int  \mathbf{Y\lm} \cdot \mathbf{Y^*\lmp}  \dif \Omega = \int  Y\lm Y^*\lmp  \dif \Omega =W \delta_{\rm l'l}\delta_{\rm m'm}
\end{equation}
\begin{equation}\label{eq.prop2a}
\int  \boldsymbol{\Psi}\lm \cdot  \boldsymbol{\Psi}^*\lmp  \dif \Omega = W l (l+1)\delta_{\rm l'l}\delta_{\rm m'm}
\end{equation}
where the superscript ``$*$'' denotes complex conjugate, $\dif \Omega = \sin\theta \dif \theta \dif \varphi$ the surface area element, $W$ a normalisation constant and $\delta$  is the Kronecker delta function. Therefore, the surface integrals in Equations (\ref{eq.prop1}) and (\ref{eq.prop2a}) are only non-null when $l=l'$ and $m=m'$. The vector spherical harmonics $\mathbf{Y\lm}$ and $\boldsymbol{\Psi}\lm$ are given by
\begin{equation}
\mathbf{Y\lm} \tp =Y\lm \tp \hat{r}
\end{equation}
and 
\begin{eqnarray}
\boldsymbol{\Psi}\lm \tp = r \boldsymbol{\nabla} {Y\lm \tp} = r \left(\frac{\partial Y\lm}{\partial \theta} \hat{\theta} + \frac{1}{\sin \theta} \frac{\partial Y\lm}{\partial \varphi} \hat{\varphi} \right)\nonumber \\
=r \left[(l+1) Z\lm \hat{\theta} + (l+1) X\lm \hat{\varphi} \right] \label{eq.temp1}
\end{eqnarray}
where Equations (\ref{eq.xlm}) and (\ref{eq.zlm}) were used in the last equality.  In the present study, given that $c\lm$ is already absorbed in the definition of $P\lm$, the normalisation constant is $W=1$, but different normalisations are adopted across different disciplines \citep{1969SoPh....9..131A}. We take $r$ to be the normalised stellar radius and adopt $r=1$ from now on. From Equations (\ref{eq.prop2a}) and (\ref{eq.temp1}), we have
\begin{equation}\label{eq.prop2}
\int (Z\lm Z^*\lmp + X\lm X^*\lmp ) \dif \Omega = \frac{W l}{ (l+1)}\delta_{\rm l'l}\delta_{\rm m'm} \, .
\end{equation}
%

\subsubsection{Derivation of $\alpha\lm$}
The coefficients $\alpha\lm$, $\beta\lm$ and $\gamma\lm$ are complex numbers, which are decomposed into their real and imaginary parts (e.g., $\alpha\lm = \Re({\alpha\lm})+ i\Im(\alpha\lm)$). Here, for completeness, we derive the equations to compute $\alpha\lm$, noting that this derivation can also be found in many textbooks.  To start, we multiply Equation (\ref{eq.br}) by $Y\lmp^*$ and integrate it over the stellar surface
\begin{eqnarray}
\int B_r\tp Y\lmp^* \dif \Omega = \int \sum\lm \alpha\lm Y\lm Y\lmp^* \dif \Omega \\ \nonumber
= \sum\lm \alpha\lm W \delta_{\rm l'l}\delta_{\rm m'm} =  \alpha\lmp W 
\end{eqnarray}
where we used Equation (\ref{eq.prop1}). Noting that $Y\lmp^* = \plmprime [\cos (m'\varphi) -i \sin(m'\varphi)]$, the real and imaginary parts of previous equation are (dropping the prime symbols)
\begin{equation}\label{eq.alphare}
\Re({\alpha\lm}) = \frac1W \int B_r\tp \plm  \cos (m\varphi) \dif \Omega \, ,
\end{equation}
\begin{equation}\label{eq.alphaim}
\Im({\alpha\lm}) = - \frac1W \int B_r\tp \plm \sin (m\varphi) \dif \Omega \, .
\end{equation}
Therefore, from an observed distribution of $ B_r\tp$, one can then solve for $\alpha\lm$ using Equations (\ref{eq.alphare}) and (\ref{eq.alphaim}).

\subsubsection{Derivation of $\beta\lm$}
The process to derive the coefficients $\beta\lm$ and $\gamma\lm$ involves a longer mathematical manipulation, since they are part of two coupled equations. The trick to derive $\beta\lm$ is to multiply Equation (\ref{eq.btheta}) by $Z^*\lmp$ and Equation (\ref{eq.bphi}) by $-X^*\lmp$ and sum the resulting equations
\begin{eqnarray}\label{eq.eq20}
[B_\theta\tp Z\lmp^* - B_\varphi\tp X^*\lmp ]  =\nonumber \\
\sum\lm \left\{ \beta\lm (Z\lmp^* Z\lm +X\lmp^* X\lm) + \gamma\lm (Z\lmp^* X\lm   - X\lmp^* Z\lm) \right\}.
\end{eqnarray}
We now integrate Equation (\ref{eq.eq20}) over the stellar surface to obtain
\begin{eqnarray}\label{eq.btheta_partial}
\int [B_\theta\tp Z\lmp^* - B_\varphi\tp X^*\lmp ] \dif \Omega =\nonumber \\
\int \sum\lm  \beta\lm (Z\lmp^* Z\lm +X\lmp^* X\lm) \dif \Omega \nonumber \\
+ \int  \sum\lm \gamma\lm (Z\lmp^* X\lm   - X\lmp^* Z\lm)\dif \Omega  
\end{eqnarray}
After some algebraic manipulation, we can demonstrate that the integral $ \int  \sum\lm \gamma\lm  (Z\lmp^* X\lm   - X\lmp^* Z\lm)\dif \Omega =0$, so that  the second term on the right hand side is null (see Appendix \ref{ap.demonstrations}). Using the orthogonal property (\ref{eq.prop2}), Equation (\ref{eq.btheta_partial}) becomes
\begin{eqnarray}
\int  [B_\theta\tp Z\lmp^* - B_\varphi\tp X^*\lmp ]  \dif \Omega =\nonumber \\
=    \sum\lm \beta\lm  \frac{W l}{ (l+1)}\delta_{\rm l'l}\delta_{\rm m'm} =   \beta\lmp \frac{W l'}{ (l'+1)} \, .
\end{eqnarray}
Substituting Equations (\ref{eq.xlm}) and (\ref{eq.zlm}) into last equation, the real and imaginary parts of $\beta\lm$ become (dropping the prime symbols)
\begin{equation}\label{eq.betare}
\Re( \beta\lm) = \frac{1}{Wl} \int \left[ B_\theta \cos (m\varphi) \dplm + B_\varphi \frac{m\sin (m\varphi)}{\sin \theta} \plm \right]  \dif \Omega \, ,
\end{equation}
\begin{equation}\label{eq.betaim}
\Im( \beta\lm) = \frac{-1}{Wl} \int \left[ B_\theta \sin(m\varphi) \dplm - B_\varphi  \frac{m\cos (m\varphi)}{\sin \theta} \plm  \right] \dif \Omega \, .
\end{equation}
%

\subsubsection{Derivation of $\gamma\lm$}
To derive $\gamma\lm$ we start by multiplying Equation (\ref{eq.btheta}) by $X^*\lmp$ and Equation (\ref{eq.bphi}) by $Z^*\lmp$ and summing the resulting equations:
\begin{eqnarray}\label{eq.eq15}
[B_\theta\tp X\lmp^* + B_\varphi\tp Z^*\lmp ]  =\nonumber \\
\sum\lm \left\{ \beta\lm (X\lmp^* Z\lm -Z\lmp^* X\lm) + \gamma\lm (X\lmp^* X\lm   + Z\lmp^* Z\lm) \right\}.
\end{eqnarray}
We now integrate Equation (\ref{eq.eq15}) over the stellar surface to obtain
\begin{eqnarray}\label{eq.bgamma_partial}
\int [B_\theta\tp X\lmp^* + B_\varphi\tp Z^*\lmp ] \dif \Omega =\nonumber \\
\int \sum\lm  \beta\lm (X\lmp^* Z\lm -Z\lmp^* X\lm)   \dif \Omega \nonumber \\
+ \int \sum\lm \gamma\lm (X\lmp^* X\lm   + Z\lmp^* Z\lm)  \dif \Omega \, .
\end{eqnarray}
As the integral $\int \sum\lm  \beta\lm  (X\lmp^* Z\lm -Z\lmp^* X\lm) \dif \Omega = 0$ (c.f.~Appendix \ref{ap.demonstrations}), the first term on the right hand side is null. Using the orthogonal property (\ref{eq.prop2}), we thus have
\begin{eqnarray}
\int [B_\theta\tp X\lmp^* + B_\varphi\tp Z^*\lmp ] \dif \Omega =\nonumber \\
=    \sum\lm \gamma\lm  \frac{W l}{ (l+1)}\delta_{\rm l'l}\delta_{\rm m'm} =   \gamma\lmp \frac{W l'}{ (l'+1)} \, .
\end{eqnarray}
Substituting Equations (\ref{eq.xlm}) and (\ref{eq.zlm}) into last equation, the real and imaginary parts of $\gamma\lm$ become (dropping the prime symbols)
\begin{equation}\label{eq.gammare}
\Re( \gamma\lm) = \frac{-1}{Wl} \int \left[B_\theta \frac{m\sin (m\varphi)}{\sin \theta} \plm - B_\varphi \cos (m\varphi) \dplm \right]  \dif \Omega \, ,
\end{equation}
\begin{equation}\label{eq.gammaim}
\Im( \gamma\lm) = \frac{-1}{Wl} \int \left[B_\theta \frac{m\cos (m\varphi)}{\sin \theta} \plm + B_\varphi \sin (m\varphi) \dplm \right]   \dif \Omega \, .
\end{equation}

We provide in Appendix \ref{ap.num_int} the discretised forms of Equations (\ref{eq.alphare}), (\ref{eq.alphaim}), (\ref{eq.betare}), (\ref{eq.betaim}), (\ref{eq.gammare}), (\ref{eq.gammaim}), to be used in numerical integrations.

\subsection{Toroidal and poloidal field components}
We can also express the vector magnetic field in terms of its poloidal and toroidal components, similarly to several stellar ZDI studies  \citep{2008MNRAS.388...80P,2009ARA&A..47..333D, 2015MNRAS.453.4301S, 2016MNRAS.455L..52V} and following the decomposition used by \citet[][Appendix III]{1946PhRv...69..106E, 1961hhs..book.....C}. The toroidal part of the field is associated with the terms with $\gamma\lm$ in Equations (\ref{eq.br}) to (\ref{eq.bphi}), i.e., the radial, meridional and azimuthal components of the toroidal field are, respectively
\begin{equation}\label{eq.brtor}
B_{{\rm tor},r} \tp =  0\, ,
\end{equation}
\begin{equation}\label{eq.bthetator}
B_{{\rm tor},\theta} \tp =   \sum\lm  \gamma\lm  \frac{ im \plm  e^{im\varphi}}{(l+1) \sin\theta}  \, ,
\end{equation}
\begin{equation}\label{eq.bphitor}
B_{{\rm tor},\varphi} \tp = \sum\lm \gamma\lm \frac{1}{l+1} \frac{\dif \plm}{\dif \theta} e^{im\varphi}   \, .
\end{equation}
Likewise, the radial, meridional and azimuthal components of the poloidal part of the field are 
\begin{equation}\label{eq.brpol}
B_{{\rm pol},r} \tp \equiv B_r \tp =    \sum\lm \alpha\lm P\lm (\cos \theta) e^{im\varphi} \, ,
\end{equation}
\begin{equation}\label{eq.bthetapol}
B_{{\rm pol},\theta} \tp =   \sum\lm \beta\lm  \frac{1}{l+1} \frac{\dif \plm}{\dif \theta} e^{im\varphi}   \, ,
\end{equation}
\begin{equation}\label{eq.bphipol}
B_{{\rm pol},\varphi} \tp =- \sum\lm \beta\lm \frac{ im \plm  e^{im\varphi}}{(l+1) \sin\theta}  \, , 
\end{equation}
such that $\mathbf{B}_{\rm pol} + \mathbf{B}_{\rm tor} = \mathbf{B}$. In the limit of a purely axisymmetric field ($m=0$), the toroidal field has only azimuthal component and the poloidal field only has radial and meridional components (i.e., it lies in meridian planes).

Using equations (\ref{eq.brpol}) to (\ref{eq.bphipol}), we  calculate the average squared poloidal component of the magnetic field $\langle B_{\rm pol}^2 \rangle = \frac{1}{4\pi}\int \sum_k B_{{\rm pol},k}^2\tp \dif \Omega$, with $k=r, \theta, \varphi$. The fraction of poloidal fields is then $f_{\rm pol}= {\langle B_{\rm pol}^2 \rangle}/{\langle B^2 \rangle}$, where $\langle B^2 \rangle = \frac{1}{4\pi}\int \sum_k B_k^2\tp \dif \Omega$. The toroidal equivalent is $\langle B_{\rm tor}^2 \rangle =\langle B^2 \rangle  - \langle B_{\rm pol}^2 \rangle $ and $f_{\rm tor} = 1 - f_{\rm pol}$.

\section{Application to solar synoptic maps of the vector magnetic field} \label{sec.application}
\begin{figure*}
	\includegraphics[height=0.187\textwidth]{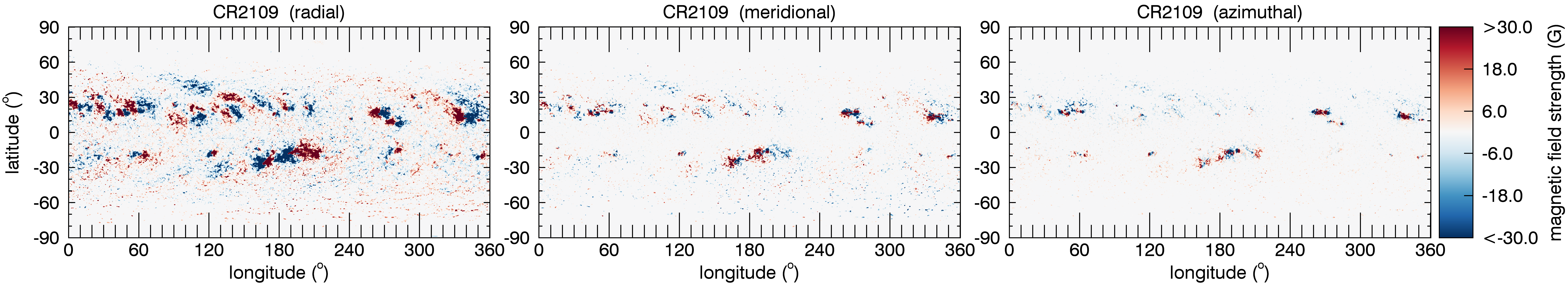}\\ \medskip
	\includegraphics[height=0.187\textwidth]{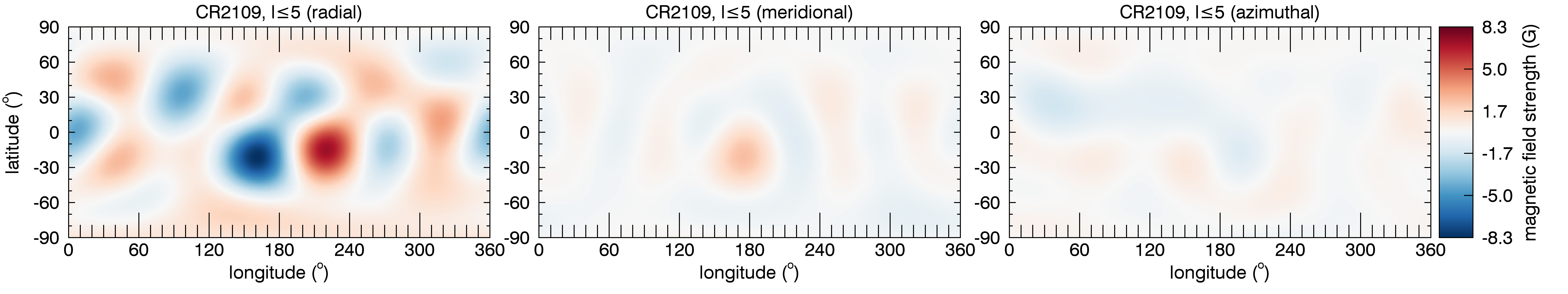}\\ \medskip
	\includegraphics[height=0.187\textwidth]{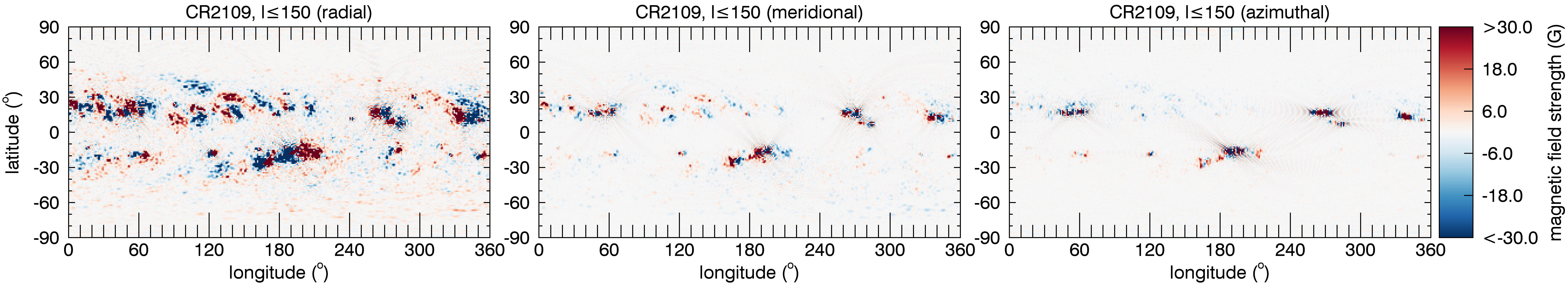}
\caption{Top row: Solar synoptic map of the vector field derived from SOLIS observations by \citet{2013ApJ...772...52G} for Carrington Rotation CR2109. 
Middle row: The large-scale magnetic field of the solar photosphere at CR2109. This field is reconstructed from the original map (upper row), by restricting the spherical harmonics reconstruction up to a degree $l\leq5$, which is typical in stellar studies.
Bottom row: The reconstructed solar magnetic field with $l \leq 150$, showing that the small-scale features that we see in the original synoptic map (top row) are recovered at $l \leq 150$.}\label{fig.sunmap}
\end{figure*}

We illustrate the application of the equations we derived in Section \ref{sec.mathematical} using a recently published synoptic map of the solar magnetic field produced with the SOLIS/VSM spectrograph \citep{2013ApJ...772...52G} and reproduced in the top row of Figure \ref{fig.sunmap}. First, we decompose each component of the magnetic field distribution using spherical harmonics and compute $\alpha\lm$, $\beta\lm$ and $\gamma\lm$ using Equations (\ref{eq.alphare}), (\ref{eq.alphaim}), (\ref{eq.betare}), (\ref{eq.betaim}), (\ref{eq.gammare}), and (\ref{eq.gammaim}). We adopt a maximum degree $l_{\rm max} = 150$ in this exercise. Table \ref{table.coeffs} shows the coefficients computed for the first five harmonics degrees of the decomposition.

\begin{table}
\caption{Spherical harmonics coefficients derived in the decomposition of the magnetic field distribution of the solar synoptic map of the vector field (top row of Figure \ref{fig.sunmap}). We adopt a maximum degree $l_{\rm max} = 150$ in the decomposition, but only present below the first five harmonics degrees, which are used to reconstruct the large-scale field of the Sun at CR2109 (middle row of Figure \ref{fig.sunmap}).} \label{table.coeffs}
\begin{center}
\begin{tabular}{rrrrrrrrrrrrr}
\hline
$l$ & $m$  & $ \Re({\alpha\lm}) $ & $\Im({\alpha\lm}) $ & $ \Re( \beta\lm) $ & $\Im( \beta\lm) $ & $\Re( \gamma\lm)  $ & $ \Im( \gamma\lm) $ \\
&& (G)& (G)& (G)& (G)& (G)& (G)\\ \hline \hline
 $    1$ & $    0$ & $    -0.288$ & $    0$ & $    -0.272$ & $    0$ & $     0.410$ & $    0 $ \\
&$    1$ & $    -1.635$ & $    -2.930$ & $     0.222$ & $    -0.467$ & $     0.043$ & $    -0.669 $ \\
 $    2$ & $    0$ & $     0.859$ & $    0$ & $     0.097$ & $    0$ & $     0.777$ & $    0 $ \\
&$    1$ & $     0.478$ & $    -0.382$ & $     0.027$ & $     0.097$ & $     0.089$ & $     0.756 $ \\
&$    2$ & $    -2.231$ & $    -2.575$ & $    -0.551$ & $    -0.133$ & $     0.448$ & $    -0.384 $ \\
 $    3$ & $    0$ & $    -0.195$ & $    0$ & $     0.194$ & $    0$ & $    -0.119$ & $    0 $ \\
&$    1$ & $    -0.565$ & $     0.501$ & $     0.628$ & $     0.393$ & $    -0.092$ & $     0.761 $ \\
&$    2$ & $     1.074$ & $     2.487$ & $    -0.275$ & $    -0.178$ & $    -0.112$ & $     0.090 $ \\
&$    3$ & $    -0.295$ & $    -2.718$ & $     0.018$ & $    -0.536$ & $     0.423$ & $    -0.529 $ \\
 $    4$ & $    0$ & $     0.094$ & $    0$ & $    -0.394$ & $    0$ & $    -0.439$ & $    0 $ \\
&$    1$ & $    -0.323$ & $     0.103$ & $    -0.160$ & $    -0.011$ & $    -0.365$ & $    -0.171 $ \\
&$    2$ & $     0.821$ & $    -0.098$ & $     0.762$ & $     0.298$ & $    -0.332$ & $    -0.176 $ \\
&$    3$ & $     0.954$ & $     4.819$ & $     0.228$ & $    -0.088$ & $     0.277$ & $     0.449 $ \\
&$    4$ & $    -5.215$ & $    -0.564$ & $    -1.093$ & $     0.013$ & $     0.537$ & $    -0.594 $ \\
 $    5$ & $    0$ & $    -0.153$ & $    0$ & $     0.335$ & $    0$ & $    -0.180$ & $    0 $ \\
&$    1$ & $     1.839$ & $     0.241$ & $    -0.505$ & $    -0.143$ & $     0.226$ & $    -0.642 $ \\
&$    2$ & $    -0.405$ & $    -3.231$ & $    -0.031$ & $    -0.219$ & $     0.088$ & $    -0.291 $ \\
&$    3$ & $    -0.198$ & $     2.383$ & $     0.426$ & $     0.292$ & $     0.340$ & $     0.575 $ \\
&$    4$ & $     0.685$ & $     2.977$ & $    -0.068$ & $     0.330$ & $    -0.094$ & $     0.610 $ \\
&$    5$ & $     0.514$ & $    -0.323$ & $    -0.177$ & $    -0.110$ & $     0.299$ & $    -1.587 $ \\
\hline
\end{tabular}
\end{center}
\end{table}

\begin{figure}
	\includegraphics[width=0.47\textwidth]{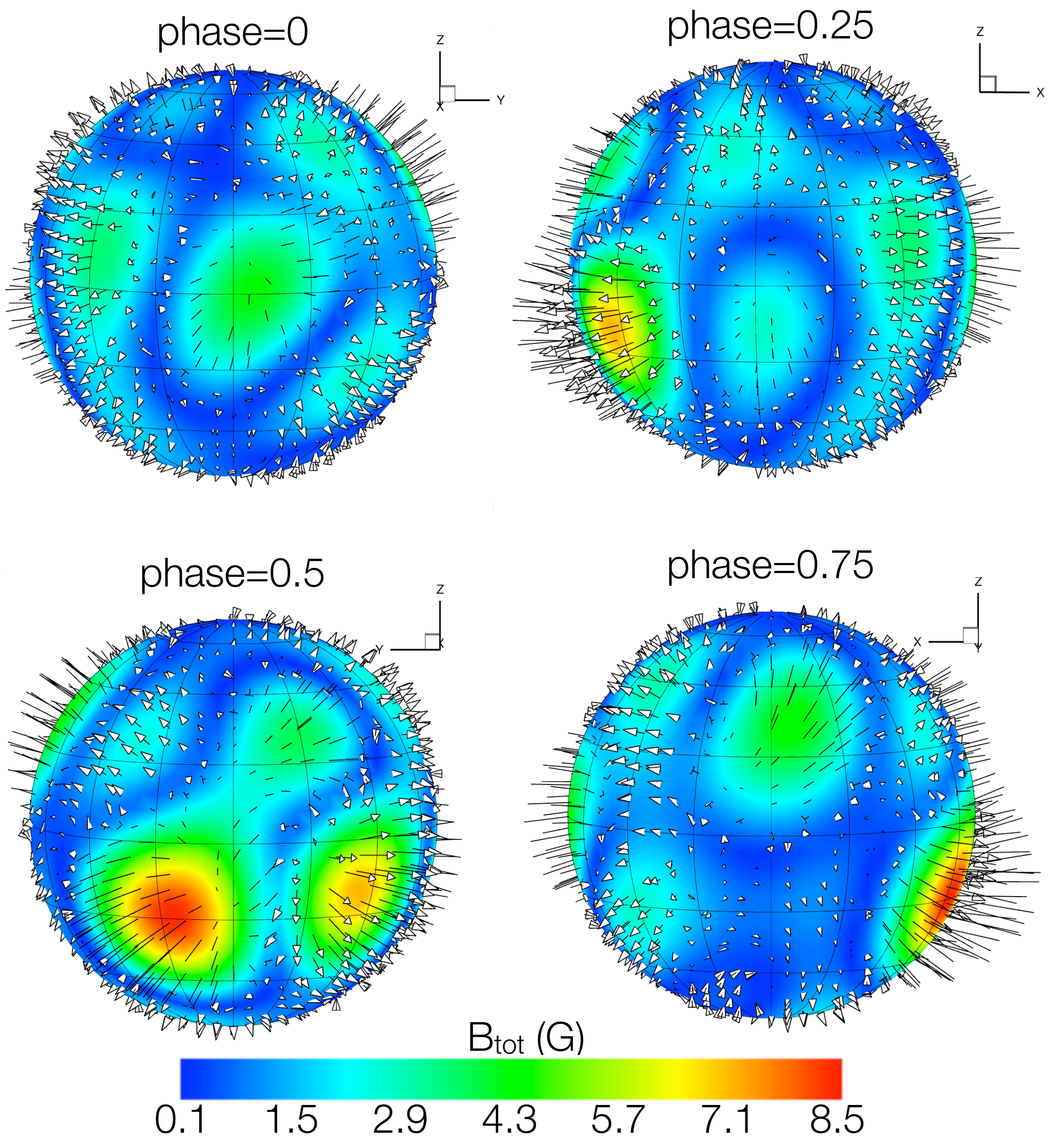}
\caption{The large-scale magnetic field vector at the solar photosphere for CR2109. The colours indicate the distribution of the magnitude of the surface magnetic field. Each panel indicates a different rotational phase.}\label{fig.vector}
\end{figure}

Using these derived coefficients, we then compute the large-scale field by restricting the sums in Equations (\ref{eq.br}) to (\ref{eq.bphi})  up to $l\leq l_{\rm max} =5$. The resulting large-scale solar vector field is shown in the second row of Figure \ref{fig.sunmap} and in Figure \ref{fig.vector}, where we overlay the magnetic vectors in the photosphere. We chose $l_{\rm max} = 5$ because this is a typical maximum degree achieved in ZDI studies of stellar magnetism.

As one goes towards larger $l_{\rm max}$ values, the differences between the original image and the reconstructed one shall become increasingly small. As a proof-of-concept, we show in the bottom row of Figure \ref{fig.sunmap} the reconstructed magnetic field with $l \leq 150$. Note that most of the small-scale features that we see in the original synoptic map (top row) are already reconstructed at this high-$l$ degree. For easier identification of these features, the contour levels of the top and bottom rows are saturated to $\pm30$~G. 

\begin{figure*}
	\includegraphics[width=\textwidth]{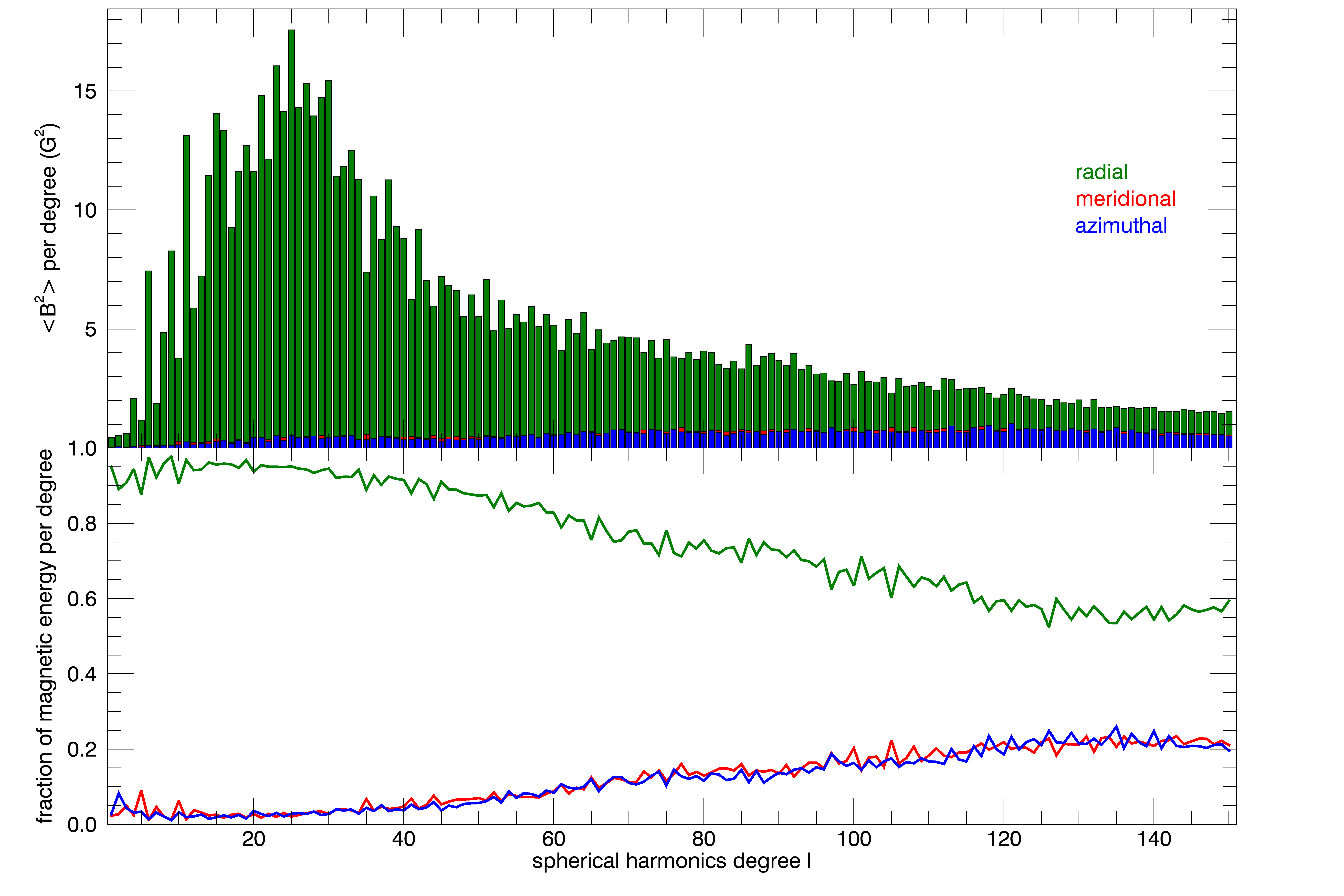}
	\caption{Top: Distribution of the magnetic energies contained in each $l$ degree for the radial (green), meridional (red) and azimuthal (blue) components, decomposed from the observed synoptic map of the vector field of the Sun (top panels of Figure \ref{fig.sunmap};  \citealt{2013ApJ...772...52G}). Bottom: the same, but for the fractional energy.}\label{fig.histogram}
\end{figure*}

Additionally, for each $l=[1, l_{\rm max}]$, we compute the magnetic field distribution of each degree using Equations (\ref{eq.br}) to (\ref{eq.bphi}), i.e., the sums are restricted to a single degree $l$, with sums over orders $|m|\leq l$. With that, we compute the average squared magnetic field (i.e., proportional to the magnetic energy) $\langle B_k^2 \rangle = \frac{1}{4\pi}\int B_k^2\tp \dif \Omega$, where $k=r, \theta$ or $\varphi$. The distributions of the average magnetic energies (radial, azimuthal and meridional) for each $l$-degree are shown in the top panel of Figure~\ref{fig.histogram}, while the bottom panel shows the fractional energy. For $l\lesssim40$, more than 90\% of the total magnetic energy at each degree is contained in the radial component. For $l\gtrsim40$, the contribution from the azimuthal and meridional components at each spherical harmonic degree increases and, at $l=150$, these components together are responsible for $\sim 45\%$ of the magnetic energy. This shows that, as we go towards small-scale fields (increasing $l$), the azimuthal and meridional energies in each degree starts to become more significant, and that at large scales, the energy is mainly concentrated in the radial component. The peak at $l\sim 25$ in the top panel of Figure~\ref{fig.histogram} is related to spatial scales of  $\sim 180^{\rm o}/l \simeq 7^{\rm o}$, which coincides to the angular size of solar active regions. The peak in the radial magnetic energy is, therefore, likely linked to the moment when active regions start to be resolved.

Computing the (cumulative) magnetic energies for all degrees up to a given $l_{\rm max}=150$, we find that about $10\%$ of the energy is in the azimuthal component, $7\%$ is in the meridional component, and $83\%$ is in the radial component. Our results demonstrate that the solar magnetic field is not purely radial. This could already be seen in the original synoptic map derived by \citet[][see also top panel of Figure \ref{fig.sunmap}]{2013ApJ...772...52G}.

\begin{figure}
	\includegraphics[width=\columnwidth]{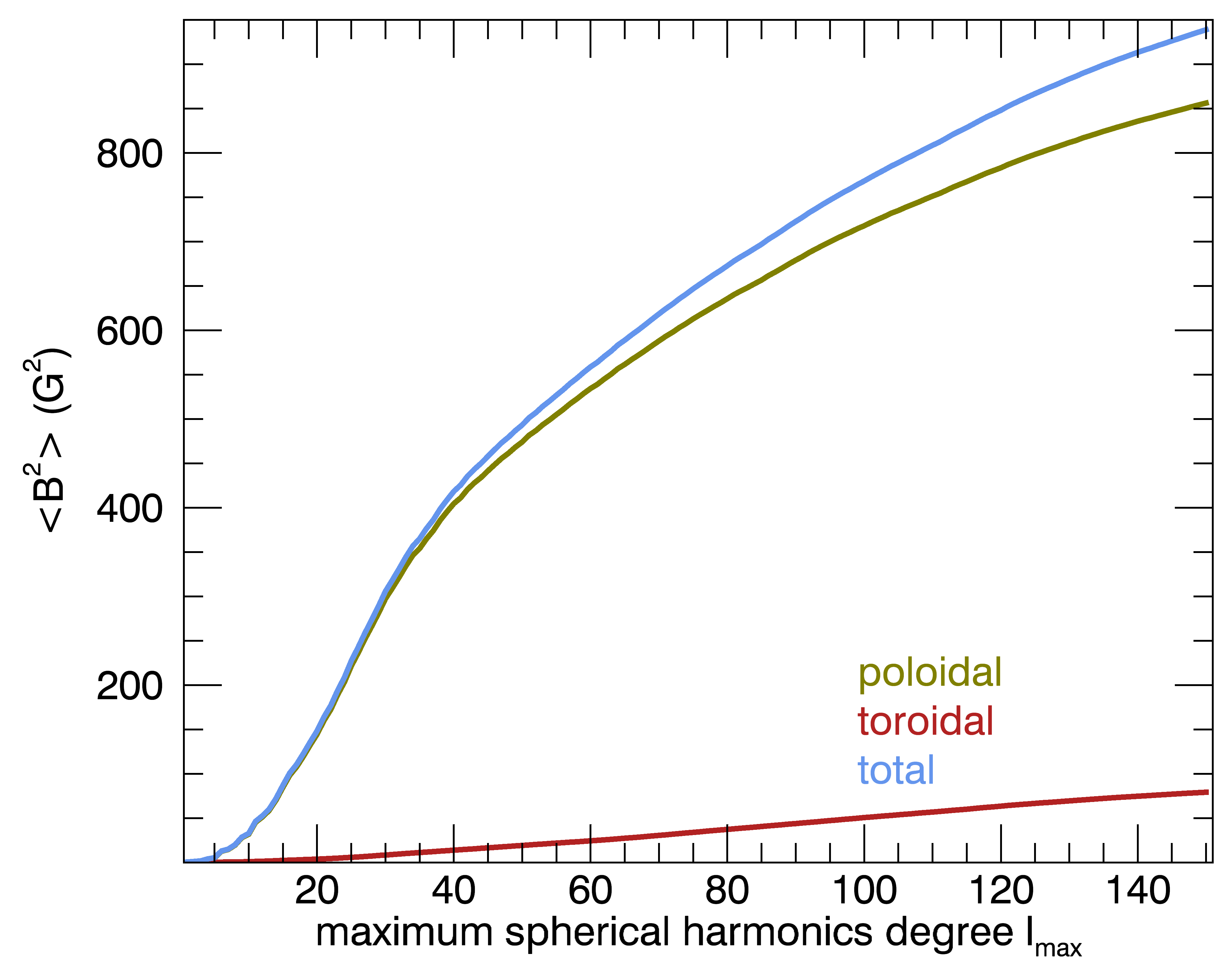}\\
	\includegraphics[width=\columnwidth]{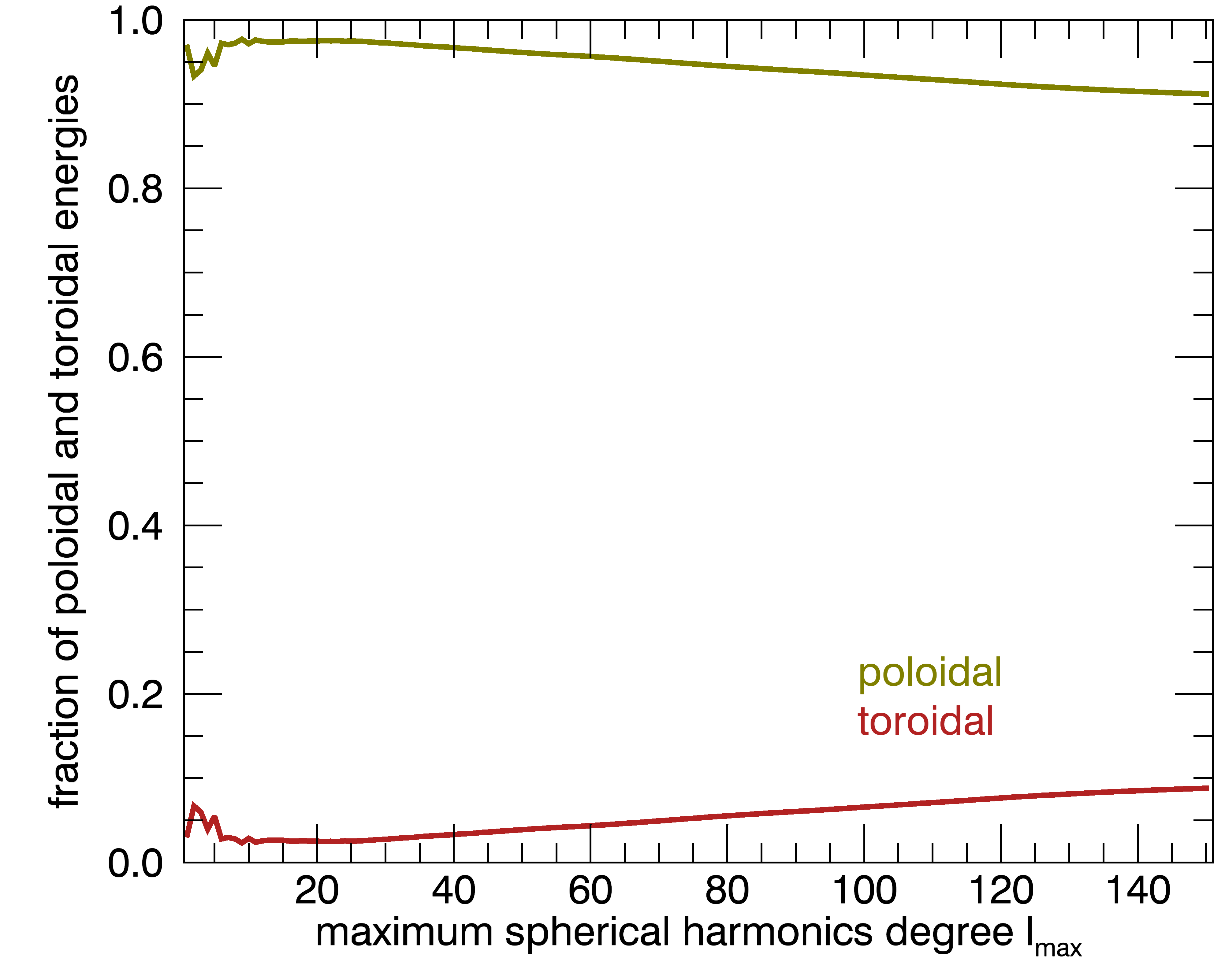}
\caption{Top: Poloidal (green), toroidal (red) and total (blue) average squared magnetic field as a function of $l_{\rm max}$. Bottom: The same, but for the fractional energies. Both panels refer to the synoptic vector field of the Sun at CR2109.}\label{fig.histogram_poltor}
\end{figure}

It is also worth converting the vector field from spherical coordinates to poloidal and toroidal components, as the latter are the ones most often quoted in papers and the ones predicted by dynamo theories. The top panel of Figure \ref{fig.histogram_poltor} shows the poloidal and toroidal energies as a function of $l_{\rm max}$. These energies were computed using Equations (\ref{eq.brtor}) to (\ref{eq.bphipol}). Since these are cumulative energies, the energies increase with $l_{\rm max}$. We also note that the dominant component of the field is poloidal: for $l_{\rm max}\lesssim 40$, more than $96\%$ of the energy is poloidal, while for $l_{\rm max}\gtrsim 40$, the poloidal energy varies monotonically from $96\%$ to $91\%$. For comparison, the map shown in Figure \ref{fig.magmap} has a similar fraction of poloidal energy ($89\%$) as the solar map at CR2109. Although the fractions of poloidal/toroidal fields are similar, the magnetic field distributions differ. Figure \ref{fig.histogram_poltor} demonstrates that the solar magnetic field during CR2109 is mainly poloidal and Figures \ref{fig.sunmap} and \ref{fig.histogram} show that it is dominated by the radial component. 

\begin{figure*}
	\includegraphics[height=0.187\textwidth]{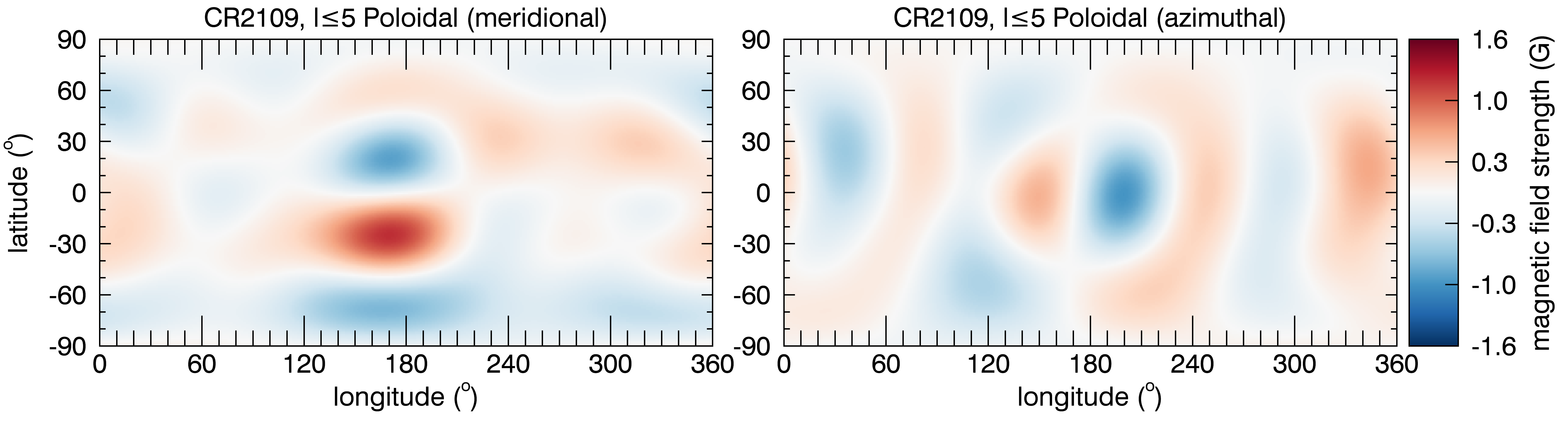}\\ \medskip
	\includegraphics[height=0.187\textwidth]{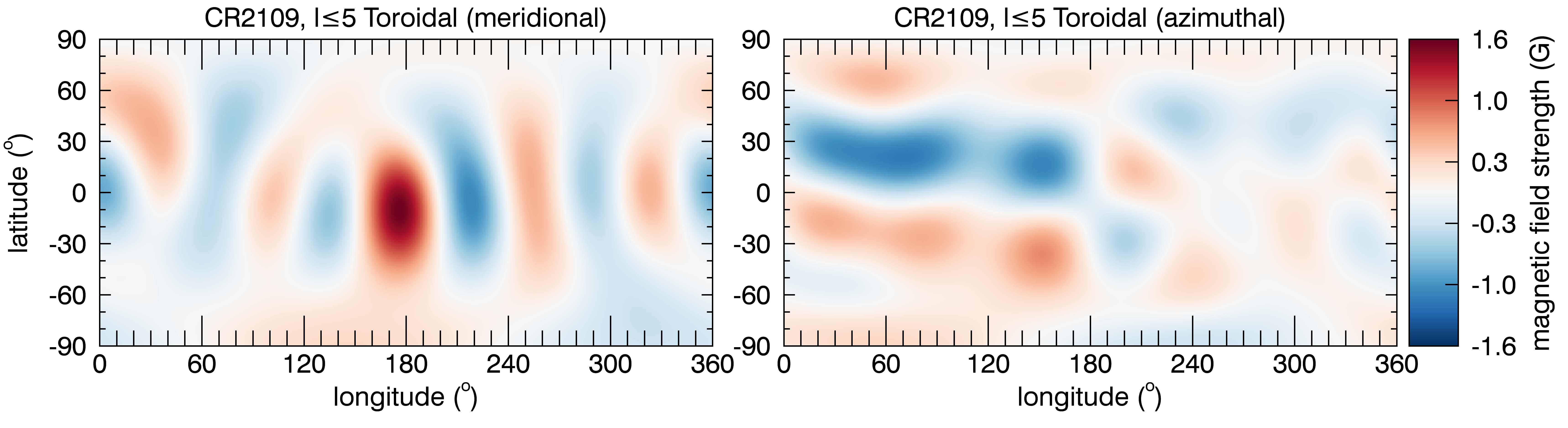}
\caption{The large-scale field of the Sun is decomposed into poloidal and toroidal vector fields, whose meridional and azimuthal components are shown here.}\label{fig.poltor}
\end{figure*}

Lastly, we show in Figure \ref{fig.poltor} the poloidal and toroidal vector fields of the large-scale field of the Sun ($l_{\rm max}=5$). Since $B_{{\rm pol},r} \equiv B_r$ and this is shown in the second row of Figure \ref{fig.sunmap}, $B_{{\rm pol},r}$ is not repeated in Figure \ref{fig.poltor}. For $l \leq 5$, $94.5\%$ of the energy is contained in the poloidal component, which is highly non-axisymmetric, with only $4.6\%$ of the energy in modes with $m=0$ or $m<l/2$.\footnote{At CR2109, the dipolar axis is at a latitude of $-7 ^{\rm o}$, i.e., it lies almost along the equatorial plane, contributing to the non-axisymmetry of the solar magnetic field.}

\section{Discussion and Conclusions}\label{sec.conclusions}
In this study, we derived a mathematical formalism to decompose any synoptic map of a vector magnetic field in terms of spherical harmonics. Given an observed bi-dimensional (longitude versus latitude) distribution of the vector magnetic field ($B_r$, $B_\theta$ and $B_\varphi$ components), we presented a self-consistent derivation of the spherical harmonics coefficients $\alpha\lm$, $\beta\lm$ and $\gamma\lm$ (Equations (\ref{eq.alphare}), (\ref{eq.alphaim}), (\ref{eq.betare}), (\ref{eq.betaim}), (\ref{eq.gammare}), and (\ref{eq.gammaim}), or, in their discretised forms, Equations (\ref{eq.alphared}) to (\ref{eq.gammaimd})).  Our description is entirely consistent with the description adopted in several ZDI studies \citep[e.g.][]{2006MNRAS.370..629D}. 

Synoptic maps of the vector field are currently available from solar \citep{2013ApJ...772...52G} and stellar \citep[e.g.][]{2006MNRAS.370..629D,2015MNRAS.453.3706D,2010MNRAS.407.2269M,2012MNRAS.423.1006F,2012A&A...540A.138M} observations and are also synthesised in numerical simulations of dynamo and magnetic flux transport studies \citep[e.g.][]{2011A&A...528A.135I,gibb2016}. Our method is particularly relevant for comparing the results of the lower resolution stellar synoptic maps against the higher resolution solar or synthetic synoptic maps of the vector field. By filtering out the spherical harmonics with high degrees, our method allows one to transform high-resolution maps into maps with similar resolution as those derived in stellar studies, allowing, therefore, their direct comparison.

In Section \ref{sec.application}, we provided an example of the use of the equations we derived by providing an application to the study of the solar magnetic fields. For that, we used a synoptic map of the vector field of the Sun reconstructed at Carrington Rotation CR2109 \citep{2013ApJ...772...52G}. We separated the large- and small-scale structure of the solar field by decomposing each component of the magnetic field distribution using spherical harmonics: we computed the spherical harmonics coefficients $\alpha\lm$, $\beta\lm$ and $\gamma\lm$ out to a maximum degree $l_{\rm max} = 150$ (see Table \ref{table.coeffs} for the coefficients corresponding to the first five $l$-degrees of the spherical harmonics decomposition). Using these coefficients, we then computed the large-scale field by restricting the sums up to $l_{\rm max}=5$ in Equations (\ref{eq.br}) to (\ref{eq.bphi}) (see Figure \ref{fig.sunmap}, second row). We have chosen $l_{\rm max}=5$ because this is a typical maximum degree achieved in ZDI studies of stellar magnetism.\footnote{We remind the reader that the larger the $l$ degree, the faster is the decay of the magnetic field with distance \citep[e.g.][]{2010RPPh...73l6901G}, such that, at a few solar/stellar radii, the dominant magnetic field components are the ones with small $l$ degrees, i.e., dipole, quadrupole, octupole, etc.}  

We showed that, at CR2109, the radial component of the solar magnetic field is the dominant one (Figure \ref{fig.histogram}), and that, for $l_{\rm max}=150$, $83\%$ of the energy is in the radial component, while $10\%$ is in the azimuthal and $7\%$ is in the meridional components. The non-radial nature of the photospheric solar magnetic field was already shown by \citet{2013ApJ...772...52G} and here we demonstrated that the large-scale field of the Sun is not purely radial either. By converting the vector field from spherical coordinates (which are easy to visualise) to poloidal and toroidal components (the ones most often quoted in papers and the ones predicted by dynamo theories), we showed that the solar magnetic energy at CR2109 is mainly ($>90\%$) poloidal (Figure \ref{fig.histogram_poltor}). 

Another way to separate large- and small-scale fields is by segregating weak and strong field regions (associated with the large- and small scale fields, respectively) using a cut-off criterion in the magnetic field strength. This has been adopted in several studies \citep[e.g.,][]{2006ApJ...646L..85Z, 2013ApJ...772...52G}, who considered the strong radial fields to have $|B_r| > 1000$~G and  weak fields to have $100~{\rm G} < |B_r| < 500~{\rm G}$. These approaches have the advantage of being straightforward and do not require long mathematical derivations such as the ones we presented here. The downside of  approaches that use a cut-off in field strength is that they do not ensure the connectivity between the three magnetic components. For example, a region with $|B_r| > 1000$~G does not necessarily have $|B_\theta|$ or $|B_\varphi| > 1000$~G; if the same cut-off threshold is used for the three components, the strong fields in the radial, meridional and azimuthal components might not be connected to the same spatial region in the star. In addition, each point in the solar photosphere is likely to be composed of fields with different scales, which together add up to the observed value. Therefore, one cannot associate regions with, e.g., $|B_r| > 1000$~G to be formed of small-scale fields solely, as an underlying large-scale field is likely to be present. For this reason, we consider the use of the method presented here  to provide a more reliable way to recover the solar large-scale magnetic field in its three components.

It is also interesting to study the variation of the observed azimuthal and meridional field components, as well as the toroidal/poloidal configurations, along the 22-yr solar cycle. Unfortunately, so far, the solar synoptic maps of the vector field are only available for Carrignton Rotations CR2109 to 2131 covering a period of less than 2 years (from 2011 March to 2012 December, \citealt{2013ApJ...772...52G}). We defer this study for the future, when large time baselines shall become available.

\section*{Acknowledgements}
The synoptic map used in Section \ref{sec.application} was acquired by SOLIS instruments operated by NISP/NSO/AURA/NSF. 
I acknowledge support from the Swiss National Science Foundation through an Ambizione Fellowship. 
I had the pleasure to discuss this project with the following people: Jean-Fran\c cois Donati, Scott Gregory, Moira Jardine, Colin Johnstone, Rosemary Mardling, Alexei Pevtsov, and Stephane Udry. I greatly appreciate their insights.

\def\aj{{AJ}}                   
\def\araa{{ARA\&A}}             
\def\apj{{ApJ}}                 
\def\apjl{{ApJ}}                
\def\apjs{{ApJS}}               
\def\ao{{Appl.~Opt.}}           
\def\apss{{Ap\&SS}}             
\def\aap{{A\&A}}                
\def\aapr{{A\&A~Rev.}}          
\def\aaps{{A\&AS}}              
\def\azh{{AZh}}                 
\def\baas{{BAAS}}               
\def\jrasc{{JRASC}}             
\def\memras{{MmRAS}}            
\def\mnras{{MNRAS}}             
\def\pra{{Phys.~Rev.~A}}        
\def\prb{{Phys.~Rev.~B}}        
\def\prc{{Phys.~Rev.~C}}        
\def\prd{{Phys.~Rev.~D}}        
\def\pre{{Phys.~Rev.~E}}        
\def\prl{{Phys.~Rev.~Lett.}}    
\def\pasp{{PASP}}               
\def\pasj{{PASJ}}               
\def\qjras{{QJRAS}}             
\def\skytel{{S\&T}}             
\def\solphys{{Sol.~Phys.}}      
\def\sovast{{Soviet~Ast.}}      
\def\ssr{{Space~Sci.~Rev.}}     
\def\zap{{ZAp}}                 
\def\nat{{Nature}}              
\def\iaucirc{{IAU~Circ.}}       
\def\aplett{{Astrophys.~Lett.}} 
\def\apspr{{Astrophys.~Space~Phys.~Res.}}   
\def\bain{{Bull.~Astron.~Inst.~Netherlands}}    
\def\fcp{{Fund.~Cosmic~Phys.}}  
\def\gca{{Geochim.~Cosmochim.~Acta}}        
\def\grl{{Geophys.~Res.~Lett.}} 
\def\jcp{{J.~Chem.~Phys.}}      
\def\jgr{{J.~Geophys.~Res.}}    
\def\jqsrt{{J.~Quant.~Spec.~Radiat.~Transf.}}   
\def\memsai{{Mem.~Soc.~Astron.~Italiana}}   
\def\nphysa{{Nucl.~Phys.~A}}    
\def\physrep{{Phys.~Rep.}}      
\def\physscr{{Phys.~Scr}}       
\def\planss{{Planet.~Space~Sci.}}           
\def\procspie{{Proc.~SPIE}}     

\let\astap=\aap
\let\apjlett=\apjl
\let\apjsupp=\apjs
\let\applopt=\ao
\let\mnrasl=\mnras

\appendix

\section{Further derivations}\label{ap.demonstrations}
In Section \ref{sec.mathematical}, we affirmed that the following integral in Eq.~(\ref{eq.btheta_partial})
\begin{eqnarray}\label{eq.dem1}
 \int  \sum\lm \gamma\lm  (Z\lmp^* X\lm   - X\lmp^* Z\lm)\dif \Omega 
 \end{eqnarray}
and the following integral in Eq.~(\ref{eq.bgamma_partial})
\begin{eqnarray}\label{eq.dem2}
 \int \sum\lm  \beta\lm  (X\lmp^* Z\lm -Z\lmp^* X\lm) \dif \Omega 
\end{eqnarray}
are both null. Here, we demonstrate these affirmations. As both demonstrations are fairly similar, we only detail the mathematical derivation for one of these integrals.

We start by rewriting the first term in Eq.~(\ref{eq.dem2}) using Eqs.~(\ref{eq.xlm}) and (\ref{eq.zlm}) 
\begin{equation}
X\lmp^* Z\lm  
=   \frac{-im' e^{i(m-m')\varphi} }{(l+1)(l'+1) \sin\theta}  P\lmp \frac{\dif \plm}{\dif \theta} 
\end{equation}
and the second term as
\begin{equation}
-Z\lmp^* X\lm= \frac{-im e^{i(m-m')\varphi}}{(l'+1)(l+1) \sin\theta} \plm \frac{\dif P\lmp}{\dif \theta} \, .
\end{equation}
Using the previous two equations,  (\ref{eq.dem2}) then becomes
\begin{eqnarray}\label{eq.dem2b}
\int  \sum\lm      \frac{- \beta\lm i e^{i(m-m')\varphi}}{(l'+1)(l+1) \sin\theta} 
(m' P\lmp \frac{\dif \plm}{\dif \theta} + m \plm \frac{\dif P\lmp}{\dif \theta})
\dif \Omega .
\end{eqnarray}
Rearranging terms and using $\dif \Omega = \sin\theta \dif \theta \dif \varphi$, we now have
\begin{eqnarray}\label{eq.dem2c}
\sum\lm \left\{\frac{- \beta\lm i}{(l'+1)(l+1) } \int  e^{i(m-m')\varphi} \dif \varphi   \right.\nonumber \\
\left. \int \frac{1}{\sin\theta} \left(m' P\lmp \frac{\dif \plm}{\dif \theta} + m \plm \frac{\dif P\lmp}{\dif \theta}\right)\sin \theta \dif \theta \right\} .
\end{eqnarray}
As $\int  e^{i(m-m')\varphi} \dif \varphi = 2 \pi \delta_{\rm m'm}$, 
the terms that contribute to the sum in (\ref{eq.dem2c}) are those when $m=m'$, and the expression (\ref{eq.dem2c})  can be rewritten as
\begin{eqnarray}\label{eq.dem2d}
\sum\lm \frac{- 2\pi \beta\lm m i}{(l'+1)(l+1) } \int  \left( P_{\rm l'm} \frac{\dif \plm}{\dif \theta} +  \plm \frac{\dif P_{\rm l'm}}{\dif \theta}\right) \dif \theta .
\end{eqnarray}
Since
\begin{equation}
\int  \left( P_{\rm l'm} \frac{\dif \plm}{\dif \theta} +  \plm \frac{\dif P_{\rm l'm}}{\dif \theta} \right) \dif \theta = \int \frac{\dif}{\dif \theta } ( P_{\rm l'm} \plm) \dif \theta = 0 \, ,
\end{equation}
this implies that (\ref{eq.dem2c}) and, consequently, (\ref{eq.dem2}), are zero.

\section{Discretised form of the equations for the spherical harmonics coefficients }\label{ap.num_int}

To obtain the spherical harmonics coefficients $\alpha\lm$, $\beta\lm$ and $\gamma\lm$, one needs to solve Equations (\ref{eq.alphare}),  (\ref{eq.alphaim}),  (\ref{eq.betare}),  (\ref{eq.betaim}),  (\ref{eq.gammare}), and  (\ref{eq.gammaim}), whose input are the observed components of the solar/stellar surface magnetic fields $B_r$, $B_\theta$ and $B_\varphi$. These quantities are often made available as bi-dimensional (latitude versus longitude), discrete arrays. It is, therefore, necessary to solve for $\alpha\lm$, $\beta\lm$ and $\gamma\lm$ in a discrete form, i.e., the integrals in Equations (\ref{eq.alphare}),  (\ref{eq.alphaim}),  (\ref{eq.betare}),  (\ref{eq.betaim}),  (\ref{eq.gammare}), and  (\ref{eq.gammaim}) become discrete sums. 

Consider arrays of $\{B_r, B_\theta, B_\varphi\}$ spaced in $n_\theta$ latitudinal grid points and $n_\varphi$ longitudinal grid points.  The discrete form of Equations (\ref{eq.alphare}),  (\ref{eq.alphaim}),  (\ref{eq.betare}),  (\ref{eq.betaim}),  (\ref{eq.gammare}), and  (\ref{eq.gammaim}) become, respectively:
\begin{equation}\label{eq.alphared}
\Re({\alpha\lm}) =  \sum_{i=1}^{n_\varphi} \sum_{j=1}^{n_\theta} B_r (\theta_j, \varphi_i) \plm(\cos \theta_j) \cos (m\varphi_i) \sin \theta_j \frac{\Delta\theta \Delta\varphi}{W} \, ,
\end{equation}
\begin{equation}\label{eq.alphaimd}
\Im({\alpha\lm}) = -\sum_{i=1}^{n_\varphi} \sum_{j=1}^{n_\theta} B_r(\theta_j, \varphi_i)  \plm(\cos \theta_j) \sin (m\varphi_i) \sin \theta_j \frac{\Delta\theta \Delta\varphi}{W} \, ,
\end{equation}
\begin{eqnarray}\label{eq.betared}
\Re( \beta\lm) = \sum_{i=1}^{n_\varphi} \sum_{j=1}^{n_\theta} \left\{ B_\theta(\theta_j, \varphi_i)  \cos (m\varphi_i) \sin \theta_j \frac{\dif \plm (\cos \theta_j)}{\dif \theta} \right. \nonumber \\ 
\left.+B_\varphi(\theta_j, \varphi_i)  {m\sin (m\varphi_i)} \plm(\cos \theta_j)  \right\} \frac{\Delta\theta \Delta\varphi}{Wl} \, ,
\end{eqnarray}
\begin{eqnarray}\label{eq.betaimd}
\Im( \beta\lm) = -\sum_{i=1}^{n_\varphi} \sum_{j=1}^{n_\theta}\left\{ B_\theta(\theta_j, \varphi_i)  \sin(m\varphi_i) \sin \theta_j \frac{\dif \plm (\cos \theta_j)}{\dif \theta} \right. \nonumber \\
\left.- B_\varphi(\theta_j, \varphi_i)   {m\cos (m\varphi_i)} \plm(\cos \theta_j)   \right\} \frac{\Delta\theta \Delta\varphi}{Wl} \, ,
\end{eqnarray}
\begin{eqnarray}\label{eq.gammared}
\Re( \gamma\lm) = - \sum_{i=1}^{n_\varphi} \sum_{j=1}^{n_\theta} \left\{B_\theta(\theta_j, \varphi_i)  {m\sin (m\varphi_i)} \plm(\cos \theta_j) \right. \nonumber \\
\left.- B_\varphi(\theta_j, \varphi_i)  \cos (m\varphi_i) \sin \theta_j \frac{\dif \plm (\cos \theta_j)}{\dif \theta}   \right\} \frac{\Delta\theta \Delta\varphi}{Wl} \, ,
\end{eqnarray}
\begin{eqnarray}\label{eq.gammaimd}
\Im( \gamma\lm) = - \sum_{i=1}^{n_\varphi} \sum_{j=1}^{n_\theta}\left\{ B_\theta(\theta_j, \varphi_i)  {m\cos (m\varphi_i)} \plm(\cos \theta_j) \right. \nonumber \\
\left.+ B_\varphi(\theta_j, \varphi_i)  \sin (m\varphi_i) \sin \theta_j \frac{\dif \plm (\cos \theta_j)}{\dif \theta}  \right\} \frac{\Delta\theta \Delta\varphi}{Wl} \, ,
\end{eqnarray}
where  $\Delta\theta = \pi/n_\theta$ and $\Delta \varphi = 2\pi/n_\varphi$. 

\bsp
\label{lastpage}
\end{document}